\documentstyle[12pt,twoside,epsfig]{article}         
\newlength{\dinwidth}     
\newlength{\dinmargin}    
\begin{document}          
\begin{titlepage}
\begin{flushright}      
{\tt LPNHE 96-01}   \\
{\tt February 1996} \\          
\end{flushright}      
\vspace*{1cm}
\begin{center}
\begin{Large}
\bf{Structure Function  Measurement  at HERA  }  
\end{Large}  
\vspace*{0.7cm}        

\begin{large}          
{Gregorio Bernardi}\\          
 LPNHE-Paris, 4 place Jussieu, 75252 Paris Cedex 05, France     
\end{large}
\end{center}

\begin{abstract}
\noindent  
Preliminary results\footnote{Invited talk given at the XXVI$^{th}$ Workshop
on High Energy Physics, Gravitation and Field Theory, held in Protvino, Russia,
June 1995. The results presented here are reflecting the status of the 
measurements reported by the H1 and ZEUS collaboration at the 1995 summer
conferences.} 
 on a  measurement 
of the proton structure function $F_2(x,Q^2)$ are reported        
for momentum transfers squared $Q^2$ between 1.5~GeV$^2$ and    
5000~GeV$^2$ and for Bjorken $x$ between $5\cdot 10^{-5}$ and $0.32$,   
using data collected by the HERA experiments H1 and ZEUS in 1994.       
$F_2$ increases significantly with decreasing $x$,   
even in the lowest reachable $Q^2$ region.   
The data are well described by a Next to Leading Order (NLO) QCD fit,   
and support within the present precision that the rise at low $x$       
within this $Q^2$ range is generated "radiatively" via the DGLAP        
evolution equations. Prospects for future 
structure function measurements at HERA are    
briefly mentioned.        
\end{abstract}    
          
          
\section{Introduction}    
The measurement of the inclusive deep inelastic lepton-proton   
scattering cross section has been of great importance for       
the understanding of the  substructure of the proton~\cite{JOEL}.        
 Experiments at HERA  extend   the previously        
accessible kinematic range up to very large squared momentum transfers, 
$Q^2 > 5 \cdot 10^4$~GeV$^2$, and to very small values of Bjorken       
$x <   10^{-4}$. 
The first observations showed a rise of the proton      
structure function $F_2(x,Q^2)$ at low $x < 10^{-2}$ with decreasing    
$x$~\cite{H1F293,ZEUSF293}, based on data collected in 1992.    
This rise was  confirmed with the more precise data of 1993     
\cite{H1F294,ZEUSF294}.   
Such a behaviour is qualitatively expected in the double leading log    
limit of Quantum Chromodynamics~\cite{ALVARO}. It is, however, not      
clarified whether the linear QCD evolution equations, as the    
conventional DGLAP evolution~\cite{DGLAP} in $\ln Q^2$ and/or the BFKL  
evolution~\cite{BFKL} in $\ln(1/x)$, describe the rise of $F_2$ or      
whether there is a significant effect due to 
      nonlinear parton recombination \cite{GLR}.     
Furthermore, it is unclear whether this rise will persist at low $Q^2$, say     
of the order of a few GeV$^2$. For example   
Regge inspired models expect $F_2$ to be     
\end{titlepage} 
flatter for small $Q^2$.  
The       
quantitative investigation of the quark-gluon interaction  dynamics at  
low $x$ is one of the major goals of  HERA. It   
requires  high precision for the     
$F_2$ measurement and  a detailed study  of the hadronic final    
state behaviour~\cite{H1BFKL}.       
          
The structure functions $F_1(x,Q^2)$, $F_2(x,Q^2)$ and $F_3(x,Q^2)$     
are related to the inclusive lepton-photon cross-section        
\begin{equation}  
  \frac{d^2\sigma^{e^+p}}{dxdQ^2} =  
       \frac{2\pi\alpha^2}{x Q^4}    
       \left[2xF_1(x,Q^2)+2(1-y)F_2(x,Q^2)   
     -\left(2y-\frac{y^2}{2}\right)xF_3(x,Q^2)\right]   
\end{equation}    
and depend on the squared four       
momentum transfer $Q^2$ and the scaling variable $x$. These variables   
are related to the inelasticity parameter $y$ and to the total squared  
centre of mass energy of the collision $s$ as $Q^2= xys$ with $s= 4     
E_e E_p$.         
However, at low $Q^2$  $xF_3$ can be neglected and the previous  
expression can be rewritten as a function of $F_2$ and $R$      
\begin{equation}  
 R \equiv \frac{F_2-2xF_1}{2xF_1} \equiv \frac{F_L}{2xF_1}.     
\end{equation}    
$R(x,Q^2)$ could not be measured yet at HERA, but can be computed,       
supposing that perturbative QCD hold, for a given set of 
parton densities. Thus $F_2(x,Q^2)$   
can be derived from the double differential cross-section $d^2\sigma/dx dQ^2$
after experimental and QED radiative corrections.    
The structure function $xF_3(x,Q^2)$ has not been measured yet, due to  
lack of statistics at high $Q^2$. However, simple differential  
cross-section $d\sigma/dQ^2$ both on neutral currents (exchange         
of a $\gamma$ or $Z^0$) or in charged current (exchange of $W^{\pm}$)   
have already been published \cite{H1CC,ZEUSCC}. In the rest of this     
paper we will consider only the case of the $\gamma$ exchange.  
          
In 1994   
the incident electron energy was $E_e = 27.5$~GeV and the       
proton energy was $E_p=820$~GeV. The data were recorded with the H1     
\cite{H1DET} and ZEUS \cite{ZEUSDET} detectors.      
A salient feature of the HERA collider       
experiments is the possibility of measuring not only the scattered      
electron but also the complete hadronic final state, apart from losses  
near the beam pipe.       
This means that the kinematic variables $x,~y$ and $Q^2$        
can be determined with complementary methods which   
experimentally are  sensitive to different systematic effects.  
The comparison  of the results obtained with different methods  
improves  the accuracy of the $F_2$ measurement. A convenient   
combination of the results ensures maximum coverage of the available    
kinematic range.  
          
In this paper after a description of the data samples (section 2) and of        
the kinematic reconstruction/event selection used (section 3) we        
provide the $F_2$ measurement in section 4. and its interpretation      
at low $Q^2$ and in terms of perturbative QCD in section 5, before      
giving some prospects in conclusion.         
          
\section{Data Samples}    
In 1994 both experiments have reduced the minimum $Q^2$ at which they   
could measure $F_2$ using several techniques.        
For DIS events at low $Q^2$  the electron is scattered under a large    
 angle    
$\theta_e$ ( the polar angles $\theta$ are defined w.r.t        
the proton beam direction, termed "forward" region). 
Therefore the acceptance of electrons in the backward region has to be  
increased or the incident electron energy to be reduced         
to go down in $Q^2$. This was realized as follows.   
          
i) both experiments were able to diminish the region around the backward        
beam pipe in which the electron could not be measured reliably in 93,   
thus increasing the maximum polar angle of the scattered electron       
(cf \cite{H1F295,ZEUSF295} for details).     
This large statistic sample, taken with the nominal HERA conditions     
is called the "nominal vertex" sample.       
Its integrated luminosity is between 2 and 3~pb$^{-1}$, depending       
on the analysis/experiment.          
          
ii) Following a pilot exercise performed last year, 
58~nb$^{-1}$ of data was collected for which the     
interaction point was shifted by +62~cm,      
in the forward direction,
resulting in an increase of the electron acceptance.        
This sample is refered to as the "shifted vertex" data sample.  
In H1 the low $Q^2$ region was also covered  by analyzing events which  
originated from the       
``early'' proton satellite bunch, present during all periods of the         
HERA operation, which     
collided with an electron bunch at a position
shifted by $+$63~cm. These data, refered       
to as the     
"satellite" data sample   
amount to $\simeq 3\%$ of the total data and correspond to a 
total "luminosity" of       
68~nb$^{-1}$.

iii) Both experiments used DIS events which underwent initial state     
photon radiation detected in an appropriate photon tagger       
to measure $F_2$ at lower $Q^2$ (so called "radiative" sample   
\cite{H1RAD}). The incident electron energy which participate in the hard  
scattering is thus reduced, and so is the $Q^2$.

The luminosity was determined from the measured cross section of        
the Bethe Heitler reaction $e^-p \rightarrow e^-p\gamma$, measuring the 
 hard photon bremsstrahlung data only.       
The precision of the luminosity for the nominal vertex position data    
amounts to 1.5\%, i.e. an improvement of a factor 3 w.r.t the analysis  
of the 1993 data. For the shifted vertex data the luminosity uncertainty
is higher (4.5\% for H1).         
 The luminosity   
of the satellite data sample was obtained from  the measured luminosity 
for the shifted vertex data multiplied by the efficiency corrected      
event ratio in a kinematic region common to both data sets.     
The uncertainty of that luminosity determination was estimated to be    
 5\%.

\section{Kinematics and Event Selection}     
          
The kinematic variables   
       of the inclusive scattering process $ep \rightarrow eX$  
can be reconstructed in different ways using measured quantities from   
the hadronic final state and from the scattered electron.       
The choice of the reconstruction method for $Q^2$ and $y$       
determines the size of systematic errors, acceptance and radiative      
corrections. The basic formulae for $Q^2$ and $y$ used in the different 
methods are summarized below, $x$ being obtained from $Q^2=xys$.        
For the electron method   
\begin{equation}  
  y_e   =1-\frac{E'_e}{E_e} \sin^{2}\frac {\theta_e} {2}        
   \hspace*{2cm}  
   Q^2_e = \frac{E^{'2}_e \sin^2{\theta_e}}{ 1-y_e}  
\end{equation}    
The resolution in $Q^2_e$ is  $4\%$ while the $y_e$  
measurement degrades as  $1/y_e$ and cannot be used  
for $y_e \leq 0.05$. In the low $y$ region it is, however,  possible    
to use the hadronic methods for which it is convenient to define the    
following variables       
\begin{equation}  
   \Sigma=\sum_h{E_h-p_{z,h}}.       
   \hspace*{2.cm} 
 p^{h~2}_T=(\sum_h{p_{x,h}})^2+(\sum_h{p_{y,h}})^2   
\end{equation}    
Here   $E,p_x,p_y,p_z$ are the four-momentum vector components  
of each particle and the summation is  over all hadronic final  
state particles.  
The standard definitions for $y_h$ and       
$\theta_h$ are    
\begin{equation}  
   y_{h}  = \frac{\Sigma}{2E_e}      
   \hspace*{3.cm} 
  \tan \frac{\theta_h}{2}=\frac{\Sigma}{p^h_T}       
\end{equation}    
The combination of $y_h$ and $Q^2_e$ defines the mixed  method  which   
is well suited for  medium and low $y$ measurements. 
The same is true for the double-angle  method which makes use   
only of $\theta_e$ and $\theta_h$ and  is thus insensitive to   
the absolute energy calibration:     
\begin{equation}  
  y_{DA} =       \frac{\tan\frac{\theta_h}{2}}       
      {\tan\frac{\theta_e}{2}+\tan\frac{\theta_h}{2}}   
   \hspace*{2cm}  
    Q^2_{DA} = 4 E_e^2 \frac{\cot\frac{\theta_e}{2}} 
      {\tan\frac{\theta_e}{2}+\tan\frac{\theta_h}{2}}   
\end{equation}    
          
The formulae for the $\Sigma$ method         
     are constructed requiring  $Q^2$ and $y$   
to be independent of the incident electron energy.   
Using the conservation of the total $E-P_z \equiv    
        \sum_i{E_i-p_{z,i}}$, the sum extending over    
all particles of the event,  $2E_e$ is replaced by   
  $  \Sigma + E'_e(1-\cos{\theta_e})$  which gives   
\begin{equation}  
   y_{\Sigma} = \frac{\Sigma}{ \Sigma + E'_e(1-\cos{\theta_e})}         
   \hspace*{2cm}  
   Q^2_{\Sigma} = \frac{E^{'2}_e \sin^2{\theta_e}}{ 1-y_{\Sigma}}       
\end{equation}    
By construction $y_{\Sigma}$ and $Q^2_{\Sigma}$ are independent of      
initial state photon radiation.      
With respect to $y_h$ the modified quantity $y_\Sigma$ is less  
sensitive to the hadronic measurement at high $y$, since the $\Sigma$   
term dominates the total $E-P_z$  of the event. At low $y$,     
$y_h$ and $y_{\Sigma}$ are equivalent.       
          
H1 measures $F_2$ with the electron and the $\Sigma$ method and after   
a complete consistency check uses the electron method for $y>0.15$      
and the $\Sigma$ method for $y<0.15$. ZEUS measures $F_2$ with the      
electron and the double angle method and for the final results  
uses the electron method at low $Q^2$ and the double angle method       
for high $Q^2$ ($Q^2 >$ 15~GeV$^2$). The binning in $x$ and $Q^2$       
was chosen to match the resolution in these variables.          
It was set at 5(8) bins per order in magnitude in x($Q^2$) for the      
H1 experiment and about twice as coarse for the ZEUS experiment in      
$x$ (similar to H1 in $Q^2$).        
          
The event selection is similar in the two experiments. Events are       
filtered on-line using calorimetric triggers which requests an  
electromagnetic cluster of at least 5~GeV not vetoed by a trigger       
element signing a beam background event. Offline, further       
electron identification criteria are applied (track-cluster link,       
shower shape and radius) and a minimum energy of 8(11)~GeV is   
requested in ZEUS(H1). H1 requests a reconstructed vertex       
within 3$\sigma$ of the expected interaction position, while ZEUS       
requires that the quantity $\delta = \Sigma +E'_e(1-\cos\theta)$        
satisfies 35~GeV $< \delta <$ 65~GeV. If no particle escapes detection, 
$\delta = 2 E =$ 55~GeV,   
so the $\delta$ cut reduces the photoproduction background and the      
size of the radiative corrections. The only significant background      
left after the selection comes from photoproduction in which    
a hadronic shower fakes an electron. In H1 for instance, 
It has been estimated      
consistently      
both from the data and from Monte Carlo simulation and amounts to       
less than 3\% except in a few bins where it can reach values    
up to 15\%. It is subtracted statistically bin by bin and an error of   
30\% is assigned to it.   
          
The acceptance and the response of the detector has been studied and    
understood in great detail by the two experiments: More         
than two millions  Monte Carlo DIS events were generated using   
DJANGO~\cite{DJANGO} and different quark distribution   
parametrizations, corresponding to an integrated luminosity of  
approximately $20$~pb$^{-1}$. The program is based on   
HERACLES~\cite{HERACLES} for the electroweak interaction        
and on  LEPTO~\cite{LEPTO} to simulate the hadronic  
final state. HERACLES includes first order radiative corrections,       
the simulation of real bremsstrahlung photons and the longitudinal      
structure function. The acceptance corrections were performed using the 
GRV~\cite{GRV}
or the MRS parametrization~\cite{MRSH}, which both 
describe rather well the HERA   
$F_2$ results of 1993 for $Q^2 > 10$~GeV$^2$.        
LEPTO uses the colour dipole model (CDM) as implemented in      
ARIADNE~\cite{CDM} which is in good agreement with data on the energy   
flow and other characteristics of the final  
state as measured by H1~\cite{H1FLOW} and ZEUS~\cite{ZEUSFLOW}. For the 
estimation of systematic errors connected with the topology     
of the hadronic final state,     the HERWIG model~\cite{HERWIG} was     
used in a dedicated analysis. Based on the GEANT program~\cite{GEANT}   
the detector response was simulated in detail.       
After this step the Monte Carlo events were subject to the same         
reconstruction and analysis chain as the real data.

\section{Structure Function Measurement}     
          
The  structure function $F_2(x,Q^2)$         
was derived after radiative corrections      
from the one-photon exchange cross section   
\begin{equation}  
  \frac{d^2\sigma}{dx dQ^2} =\frac{2\pi\alpha^2}{Q^4x}          
    (2-2y+\frac{y^2}{1+R}) F_2(x,Q^2)        
\label{dsigma}    
\end{equation}    
Effects due to $Z$ boson exchange are smaller than 2\%.         
The structure function ratio $R=F_2/2xF_1 - 1 $      
has not been measured yet at HERA and        
was calculated using the  QCD relation~\cite{ALTMAR} and the    
GRV structure function parametrization.      
Compared to the 1993 data analyses~\cite{H1F294, ZEUSF294}      
the $F_2$ measurement has been extended to lower     
$Q^2$ (from $4.5$~GeV$^2$ to  $1.5$~GeV$^2$),        
and to lower and higher $x$ (from $1.8 \cdot 10^{-4}-0.13$ to   
$5 \cdot 10^{-5}-0.32$).  
The determination of the structure function requires 
the measured event numbers to be converted   
to the bin averaged cross section    
based on the Monte Carlo acceptance calculation.     
 All detector efficiencies were determined from the data        
using the redundancy of the apparatus. Apart from very small    
extra corrections all efficiencies are correctly     
reproduced by the Monte Carlo simulation.    
The bin averaged cross section was   
corrected for higher order QED radiative contributions and a    
bin size correction was performed. This determined   
the one-photon exchange cross section        
which according to eq.\ref{dsigma}   
led to the values for $F_2(x,Q^2)$.  
          
Due to the different data sets available: "nominal vertex" data,        
 "radiative events",  "shifted vertex" data and "satellite" data,       
which have different acceptances and use     
for  a given $Q^2, x$ point different parts of the detectors, cross     
checks could be made in kinematic regions of overlap.   
The results were found to be in  very good  agreement   
with each other for all kinematic reconstruction methods used.  
          
The large available statistics  allows to make       
very detailed studies on the         
detector response: efficiencies and calibration. As a result the        
 systematic errors on many effects are reduced, compared to     
the 1993 data analysis.   
Here only a brief summary of these preliminary errors is given, 
refering the       
reader to the original and forthcoming $F_2$ publications: For the      
electron method the main source of error are the energy calibration     
(known at 1.5\% level in 1994), the knowledge of the electron   
identification efficiency and to a lesser extent the error on the polar 
angle of the scattered electron, ($\delta\theta=$1mrad) in      
particular at the lowest $Q^2$ and the radiative corrections at         
low $x$.  
For the $\Sigma$ method, the knowledge of the absolute energy scale     
for the hadrons, the fraction of hadrons which stay undetected  
in particular at low $x$, due to calorimetric thresholds        
and to a lesser extent the electron energy calibration are the  
dominating factors. For the double angle method, the major problem     
comes from the precision in the resolution of the hadronic angle        
at low $x$ and low $Q^2$. Further uncertainties common to all methods   
(selection, structure function dependance etc.) were also studied.      
The preliminary error on the 1994 data ranges between 10 and 20\%       
with expected final values for publication below 10\%.          
          
\begin{figure}[htbp]       
\begin{center}    
\epsfig{file=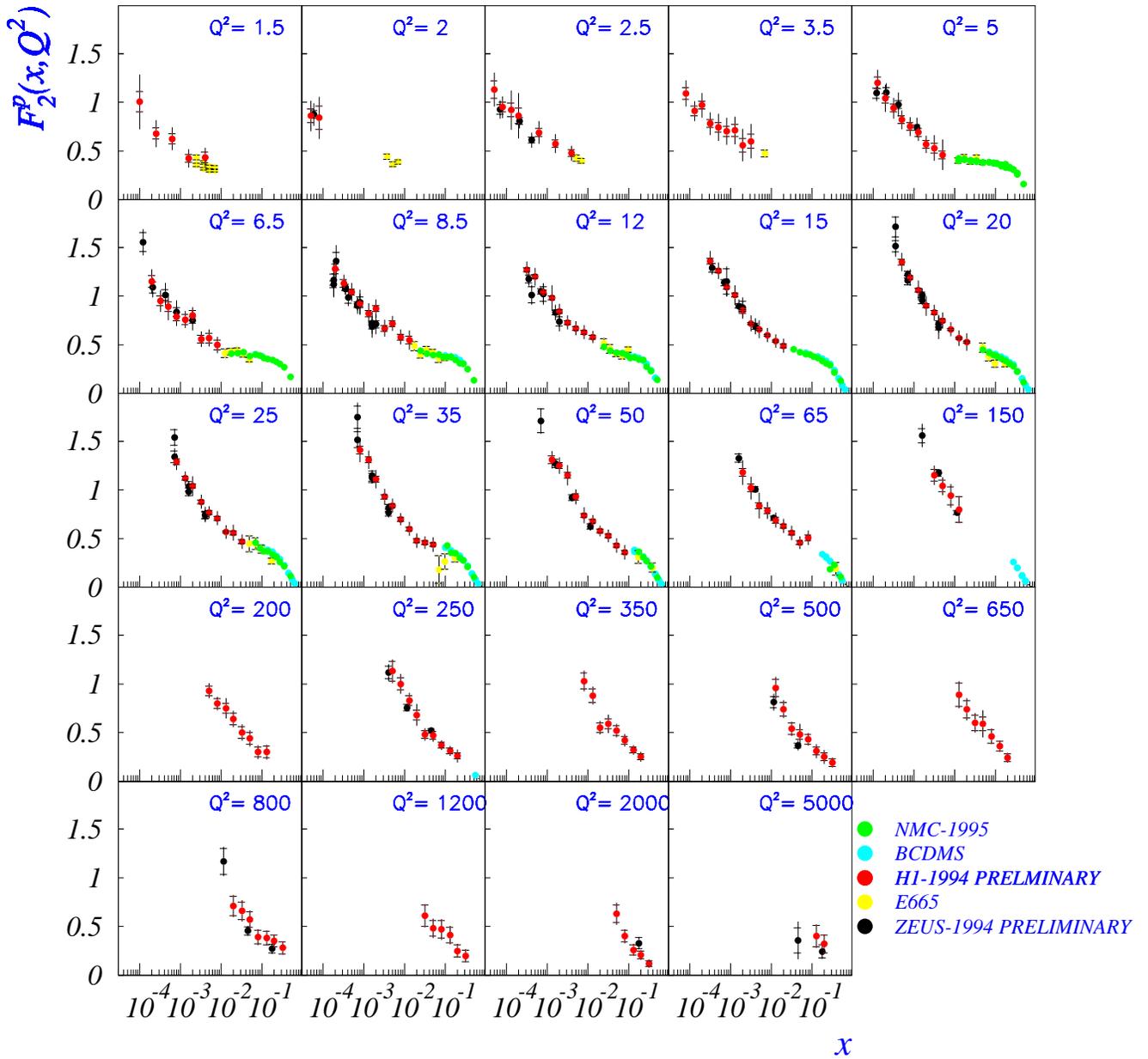,width=17.5cm,%
 bbllx=15pt,bblly=140pt,bburx=600pt,bbury=690pt}     
\end{center}      
\caption[]{\label{f2x}    
\sl Preliminary measurement of the proton structure function $F_2(x,Q^2)$   
    as function of $x$ in different bins of $Q^2$. The          
    inner error bar is the statistical error. The full error    
    represents the statistical and systematic errors added      
    in quadrature.}       
\end{figure}

The preliminary results of H1 and ZEUS are shown in fig.~\ref{f2x}, in bins     
of fixed $Q^2$. The rise of $F_2$ at low $x$ is confirmed with the      
higher precision, and is now observed down to the lowest $Q^2$  
measured (1.5~GeV$^2$). The observed good agreement between H1 and ZEUS 
and the smooth transition between the HERA and the fixed target (E665,  
NMC) data consolidates this result which can thus be confronted to      
theoretical expectations. The steepness of the low $x$ rise increases   
visibly with the $Q^2$, a characteristic expected from perturbative     
QCD. This rise cannot be attributed to the presence of          
"diffractive" events in the DIS sample since their proportion has       
been shown to stay essentially constant (~10\%) independently   
of $x$ and $Q^2$ \cite{H1DIFF,ZEUSDIFF}.     
          
\section{Low $Q^2$ and Perturbative QCD}     
          
\begin{figure}[htb]       
\begin{center}    
\epsfig{file=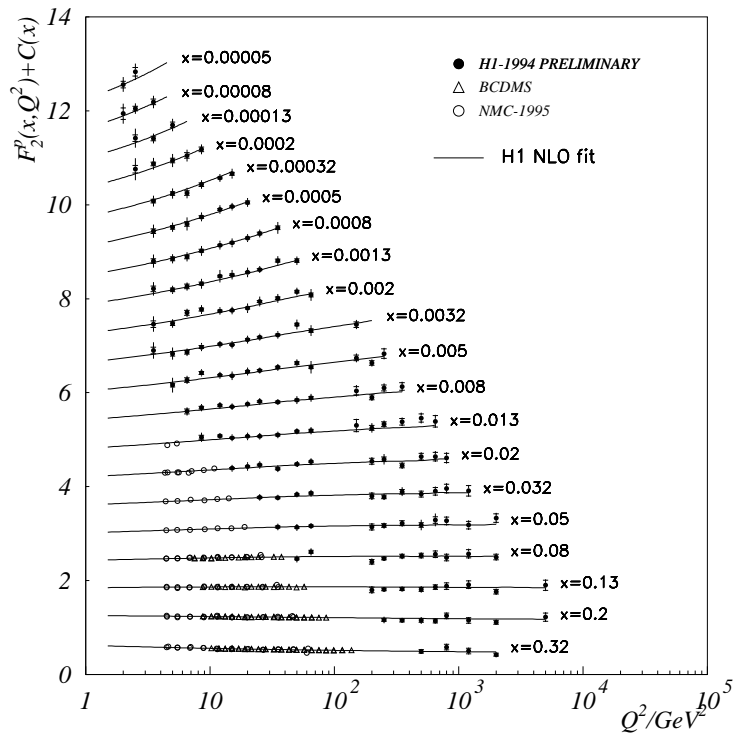,height=9.5cm,%
 bbllx=15pt,bblly=150pt,bburx=570pt,bbury=695pt}     
\end{center}      
\caption[]{\label{f2q}    
\sl Preliminary measurement of  $F_2(x,Q^2)$.      
The H1 data is consistent with the fixed target experiments     
BCDMS and NMC. The curve represents the NLO QCD-fit decribed in the text.   
The gap visible around 100 GeV$^2$ corresponds to a boundary region     
between two calorimeters in the H1 detector, which is not completely    
analyzed yet.}    
\end{figure}

A test of perturbative QCD is displayed in fig.~\ref{f2q} which represents      
the results of Next to Leading Order (NLO) QCD fit performed by H1      
as explained in \cite{H1QCD} on the data with $Q^2 \le $ 5~GeV$^2$.     
In order to constrain the structure function $F_2$ at high $x$,         
data from the fixed       
target scattering experiment NMC \cite{NMC} and BCDMS \cite{BCDMS}      
are used, avoiding        
regions where higher twist and target mass effects could become         
important.        
          
The $F_2$ behaviour can be well described by the DGLAP          
evolution equations within the present preliminary errors. The data     
for $Q^2$ values below 5~GeV$^2$ are also    
compatible with the extrapolation of the fit in this region,    
as can be seen in the figure.        
          
\begin{figure}[htb]       
\begin{center}    
\epsfig{file=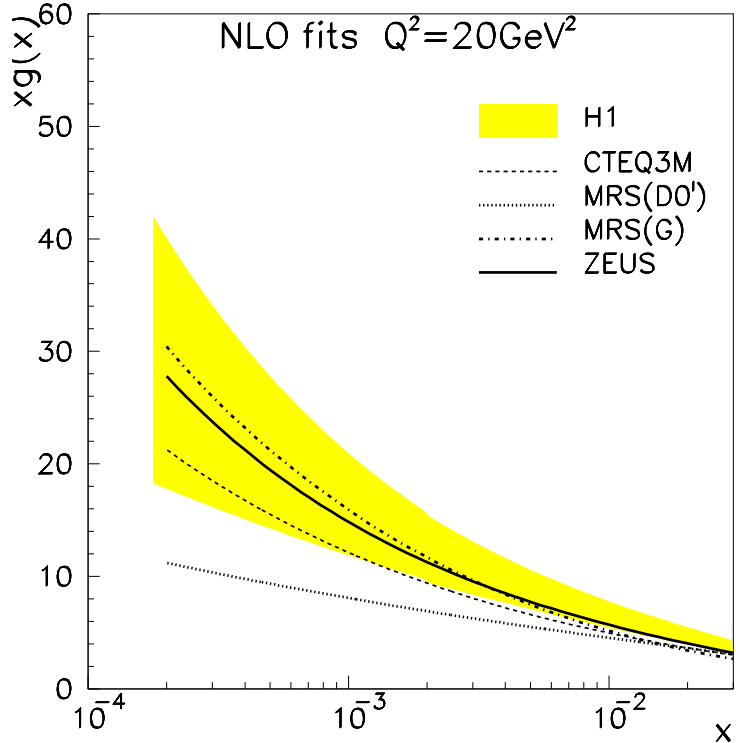,height=10cm,%
 bbllx=90pt,bblly=220pt,bburx=500pt,bbury=640pt}     
\end{center}      
\caption[]{\label{glu}    
\sl Preliminary measurement of the gluon density from NLO QCD   
    fit. The H1 result is shown with the complet systematic     
    error band, for ZEUS only the value of the gluon density is shown,  
    the error band is similar to the H1 one.}        
\end{figure}

This preliminary result is consistent with the published QCD analysis   
of H1 and ZEUS 1993 data, which allowed to determine the gluon  
density in the proton, and observe its steep rise at low $x$ as         
displayed in fig.~\ref{glu} for $Q^2=$ 20~GeV$^2$.   
          
Another test of perturbative QCD lies in observing the asymptotic    
behaviour as suggested by early studies \cite{ALVARO}.          
Ball and Forte ~\cite{BALL} have recently shown      
that evolving a flat input distribution at some $Q_0$, of the order     
of 1~GeV$^2$,     
with the DGLAP equations  
 leads to a strong rise of $F_2$ at low $x$  
in the region measured by HERA. An interesting feature is that if QCD   
evolution is the underlying dynamics of the rise, perturbative QCD predicts     
that at large $Q^2$ and small $x$    
the structure function    
 exhibits double scaling in the two variables $\sigma$ and $\rho$ defined
as: 
\begin{equation}  
\sigma \equiv \sqrt{\log(x_0/x)\cdot \log(t/t_0)}, \ \ \        
 \rho \equiv \sqrt{\frac{\log(x_0/x)}{\log(t/t_0)}} \ \ \
  \mbox{with} \ \ t\equiv \log(Q^2/\Lambda^2)           
\end{equation}

\begin{figure}[htb]       
\begin{center}    
\epsfig{file=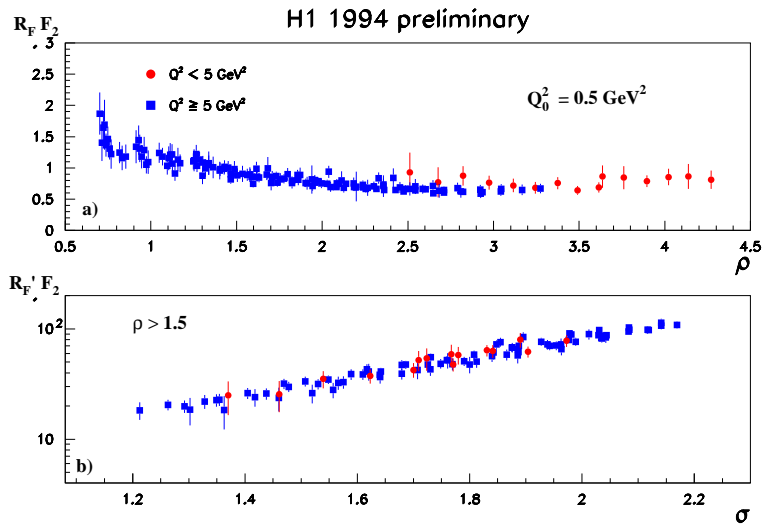,%
   height=7cm,bbllx=20pt,bblly=230pt,bburx=530pt,bbury=590pt}   
\end{center}      
\caption[]{\label{ball05}    
\sl  The rescaled structure functions        
  $R_F F_2$ and $R'_FF_2$  (preliminary) plotted versus the variables   
  $\rho$ and  $\sigma$ defined in the text. Only data with $\rho>1.5$ are       
  shown in b.}    
\end{figure}

In figure~\ref{ball05} the H1 data are presented in the         
variables $\sigma$ and $\rho$, taking        
the boundary conditions to be $x_0=0.1$ and  
$Q^2_0=0.5~{\rm GeV}^2$,  and $\Lambda^{(4)}_{\rm LO}=185~{\rm MeV}$.   
In a previous analysis \cite{H1QCD} the value  $Q^2_0=1~{\rm GeV}^2$ was
chosen,     
but the new low $Q^2$ data seems to indicate that $F_2$ is not yet flat 
for this $Q^2$ value.               
The measured values of $F_2$ are rescaled by 
\begin{equation}  
R'_F(\sigma,\rho) = 8.1\: \exp \left(\delta \frac{\sigma}{\rho}+\frac{1}{2}     
\log(\sigma) +    
\log (\frac{\rho}{\gamma})\right),   
\end{equation}    
to remove the part of the leading subasymptotic behaviour which can be  
calculated in a model independent way;       
$\log(R'_F F_2)$ is then predicted to rise linearly with $\sigma$.      
Scaling in $\rho$  can  be shown by multiplying $F_2$   
by the factor $R_F\equiv R'_Fe^{- 2\gamma \sigma}$.  
Here $\gamma \equiv 2\sqrt{3/b_0}$ with      
$b_0$ being the leading order coefficient of the $\beta$ function       
of the QCD renormalization group equation for four flavours,    
$\delta = 1.36$ for four flavours and three colours. 
          
Fig.~\ref{ball05}a shows $R_FF_2$ versus $\rho$.     
Scaling roughly sets in for          
$\rho \ge 1.5$.   
Fig.~\ref{ball05}b, for $\rho \ge 1.5$, shows  
scaling behaviour, namely a linear of        
$\log(R'_F F_2)$  with $\sigma$.     

\begin{figure}[htbp]       
\begin{center}    
\epsfig{file=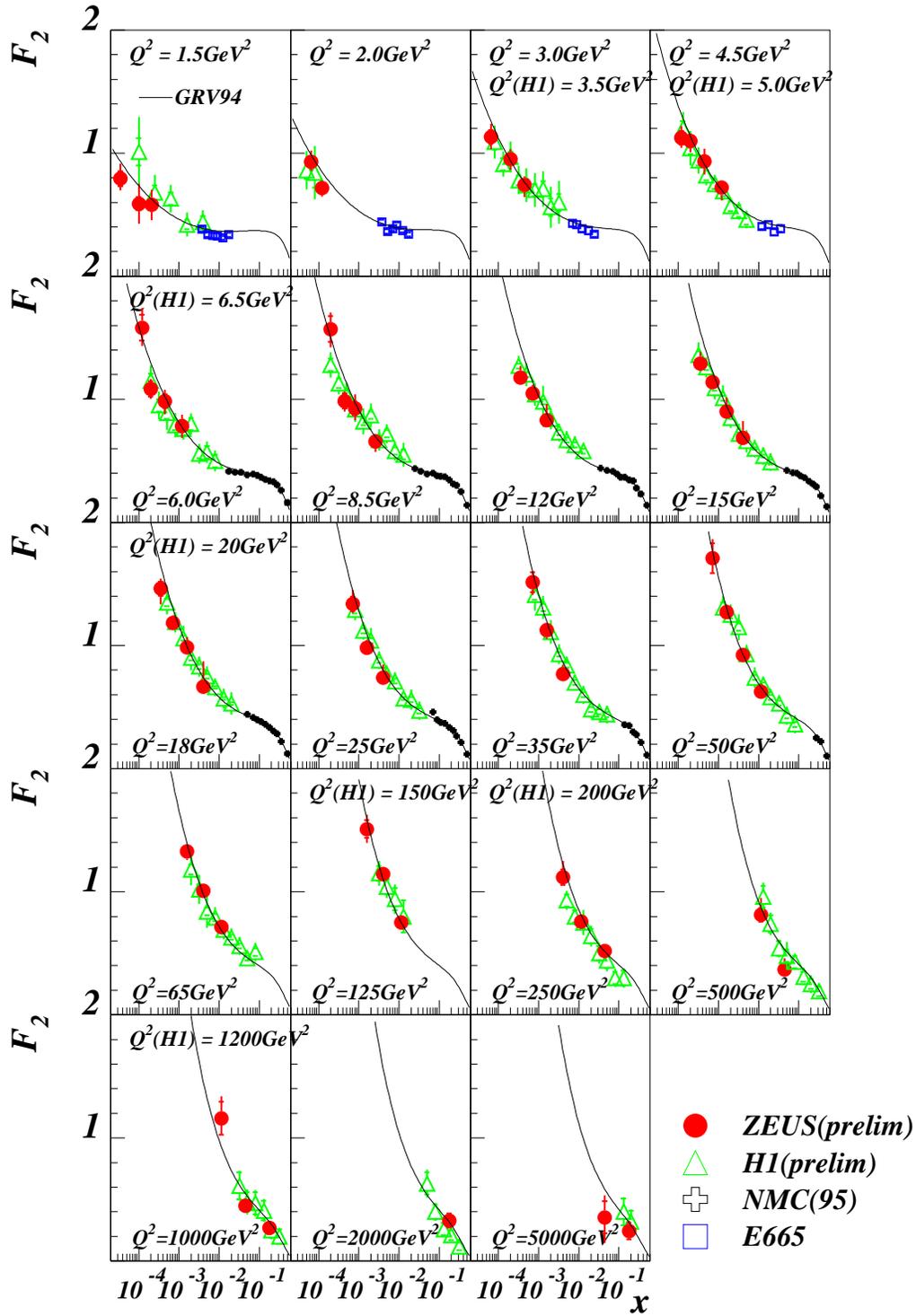,width=13.5cm,%
 bbllx=40pt,bblly=50pt,bburx=545pt,bbury=780pt}      
\end{center}      
\caption[]{\label{f2grv}  
\sl Preliminary measurement of the proton structure function $F_2(x,Q^2)$ 
by H1 and ZEUS, compared to the  
 results of  the E665 and NMC experiments and to the ``prediction'' of the 
 GRV model \cite{GRV} over the full $Q^2$ range.}       
\end{figure}

These observations suggest that perturbative QCD could be already       
valid at $Q^2=$ 1 or 2~GeV$^2$. Indeed within the present       
precision we can observe in fig.~\ref{f2grv} the validity of the        
Gl\"uck, Reya, Vogt (GRV) model \cite{GRV} which assumes that all low   
$x$ partons are generated "radiatively" starting from a very low        
initial $Q^2 =$ 0.34~GeV$^2$ scale, in which both gluon and quark       
densities are "valence" like. This result appears surprising since      
perturbative QCD does not apply at such a low scale, but the HERA       
results and the E665 \cite{E665} preliminary results follow the GRV     
expectations as early as 0.8~GeV$^2$. More precise data are needed to   
further constrain the model and draw definite conclusions       
on the dynamics underlying the low $x$ rise. Nevertheless these results 
appear already very promising for the DGLAP evolution equations         
which might not need to be supplemented by the BFKL         
evolution at low $x$, in the HERA kinematic domain.

\begin{figure}[hb]       
\begin{center}    
\epsfig{file=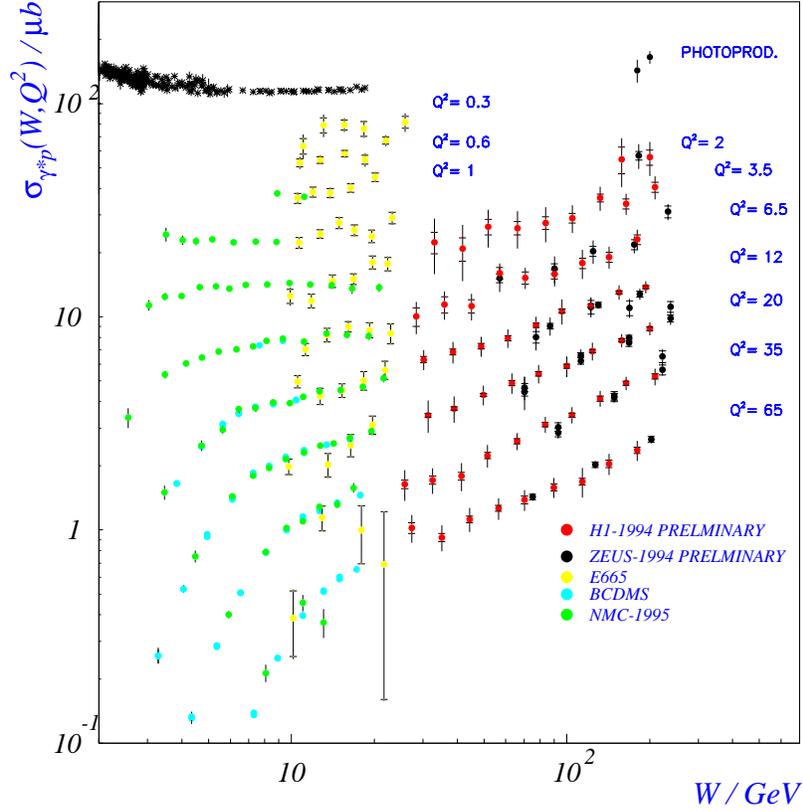,height=11cm,%
 bbllx=10pt,bblly=130pt,bburx=550pt,bbury=690pt}     
\end{center}      
\caption[]{\label{gp}     
\sl Preliminary measurement of the cross-section     
    $\sigma_{tot}(\gamma^{\ast} p)$ as a function of $W$ and $Q^2$.     
    Results from DIS are compared with the measurements in photoproduction.  
    For the readability of the plot, not all $Q^2$ bins are shown.}       
\end{figure}

Focusing now on low $Q^2$, the persistent rise of $F_2$         
at low $x$, when going down in $Q^2$         
indicates that the photoproduction  regime has not been         
reached yet.      
This can be seen in fig.~\ref{gp} which display the behaviour of the total   
cross-section of the proton-virtual photon system as a function of      
$W$, the invariant mass of the $\gamma^* p$ system   
(at low $x$, $W  \simeq \sqrt{Q^2/x}$). $F_2$ is     
related to  the total cross-section of the proton-virtual photon        
interaction ($\sigma_{tot}(\gamma^* p)) $ via        
\begin{equation}  
 \sigma_{tot}(\gamma^* p) \simeq \frac{4~\pi^2 \alpha}{Q^2} F_2(W,Q^2). 
\end{equation}    
The $\sigma_{tot}$  growth   can be contrasted with the weak rise with  
$W$   of the total real photoproduction cross-section   
   in the same range of $W$ : 20-250~GeV     
\cite{H1STOT,ZEUSSTOT}    
          
\begin{figure}[htb]       
\begin{center}    
\epsfig{file=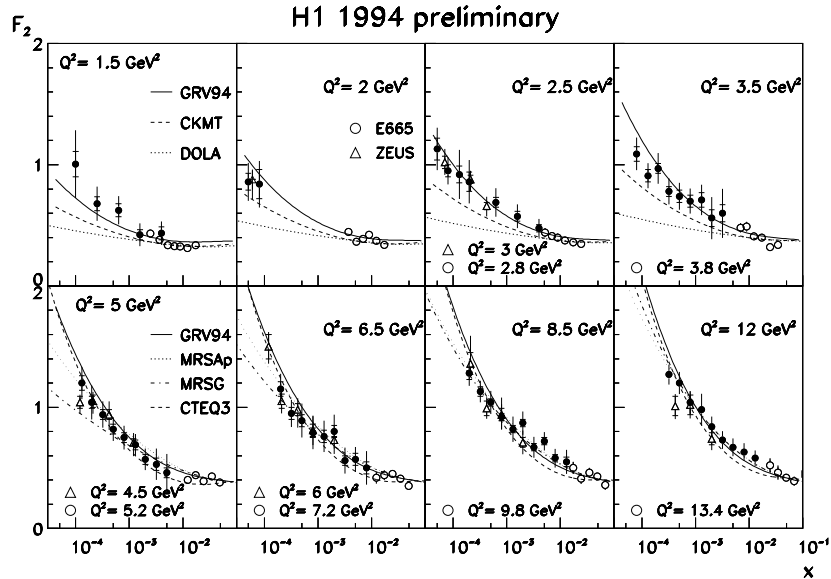,height=11.cm,angle=270.,%
 bbllx=80pt,bblly=90pt,bburx=535pt,bbury=745pt}      
\end{center}      
\caption[]{\label{lowq2}  
\sl Preliminary measurement of the proton structure function $F_2(x,Q^2)$
    in the low $Q^2$ region by H1 and ZEUS,  
    together with results from the   
    E665 experiment. Different predictions for $F_2$ are compared to the    
    data. The DOLA and CKMT curves are only shown for the upper row of $Q^2$   
    bins; CTEQ3M, MRSG and MRSA' are shown for the lower row;   
    GRV is shown for the full range.}        
\end{figure}

The Regge inspired models DOLA~\cite{DOLA}     
and CKMT\cite{CTKM} which can describe       
the behaviour of $\sigma_{tot}(\gamma p)$ predicts a rather flat        
behaviour of $F_2$ at a few GeV$^2$. As shown in fig.~\ref{lowq2},      
the DOLA model clearly fails before 1.5~GeV$^2$, while the CKMT model which    
assumes that the "bare" pomeron visible at high $Q^2$  has a higher     
trajectory intercept ($\sim 0.24$) than the "effective" pomeron involved in   
"soft" interactions ($\sim 0.08$) undershoot the data in a less critical      
manner. In this same plot we can also notice the similarity above       
5~GeV$^2$ between the different parametrizations (GRV, MRS, CTEQ        
\cite{GRV, MRSA, CTEQ} )    
which use essentially the same data to determine their parton   
distributions at the reference scale.        
          
\section{Prospects for Structure function measurements at HERA}         
          
The HERA structure function program is still in its infancy, but has    
already provided exciting results at low $x$. 
The dynamics underlying the behaviour of the        
structure function can be studied in an exclusive way   
using jets or  particle spectra,    
since HERA is a collider equipped with (two) 4$\pi$ detectors.    
In the next 2 or 3 years $F_{L}$ will be measured, by taking data at    
different beam energies in order to keep $x$ and $Q^2$ constant while   
varying $y$, thus improving the knowledge on $F_2$ and on QCD.  
The statistics will increase in such a way that a first measurement     
of $xF_3$ will be made, and a precise determination of $\alpha_s$       
should be possible. In 95, both experiments have upgraded their     
detector in the backward area in order to reach lower $Q^2$ ($\simeq$   
0.1~GeV$^2$) with good precision. This year will thus be devoted to     
understand the questions raised in this paper concerning the low $x$    
and low $Q^2$ dynamics and to open up further stringent test of         
QCD, in particular about the behaviour of the high parton       
density.  
          
\vspace*{1.cm}    
          
{\bf Acknowledgements}    
\normalsize       
          
\noindent         
I would like to thank the organizers and in particular Vladimir        
Petrov to have made such a nice workshop in the quiet town of  
Protvino, and to have invited me to discover Russia for the    
first time. I would also like to thank my close collaborators,
Ursula Bassler, Beatriz Gonzalez-Pineiro, all the friends of the
H1 structure function group and the ZEUS collaboration
with whom we obtained the results
described above. Special thanks go to Ursula             
for her  help in the finalization of this         
paper.

     
          

\begin{thebibliography}{99}          
\bibitem{JOEL}    
  for a recent review see:   
  J. Feltesse, DAPNIA-SPP-94-35(1994), Invited talk at the 27.  
  International Conference on  High Energy Physics, Glasgow, Scotland,  
  1994.   
\bibitem{H1F293}  
  H1 Collab., I. Abt et al., Nucl. Phys. {\bf B407} (1993) 515.         
\bibitem{ZEUSF293}        
  ZEUS Collab., M. Derrick et al., Phys. Lett. {\bf B316} (1993) 412.   
\bibitem{H1F294}  
  H1 Collab., T. Ahmed et al., Nucl. Phys. {\bf B439} (1995) 471.       
\bibitem{ZEUSF294}        
  ZEUS Collab., M. Derrick et al., Z. Phys. {\bf C65} (1995), 379.      
\bibitem{ALVARO}  
  A. De R\'{u}jula et al, Phys. Rev. {\bf D10} (1974) 1649.     
\bibitem{DGLAP}   
  Yu. L. Dokshitzer, Sov. Phys. JETP {\bf 46} (1977) 641; \\    
  V. N. Gribov and L.N. Lipatov, Sov. J. Nucl. Phys. 
  {\bf 15} (1972) 438 and 675; \\    
  G. Altarelli and G. Parisi, Nucl. Phys. {\bf B126} (1977) 297.        
\bibitem{BFKL}    
  E. A. Kuraev, L. N. Lipatov and V. S. Fadin, Sov. Phys. JETP {\bf 45} 
  (1977) 19; 9; \\        
  Y. Y. Bal\u{i}tsky and L.N. Lipatov, Sov. J. Nucl. Phys. {\bf 28}     
  (1978) 822.     
\bibitem{GLR}     
  L. V. Gribov, E. M. Levin and M. G. Ryskin,        
  Phys. Rep. {\bf 100} (1983) 1; \\  
  A. H. Mueller and N. Quiu, Nucl. Phys. {\bf B268} (1986) 427.         
\bibitem{H1BFKL}  
  H1 Collab, S. Aid et al., DESY preprint 95-108 (1995).        
\bibitem{H1CC}    
  H1 Collab, S. Aid et al., DESY-preprint 95-102 (1995).        
\bibitem{ZEUSCC}  
  ZEUS Collab, M. Derrick et al.,    
  Phys. Rev. Lett. {\bf 75} (1995) 1006.     
\bibitem{H1DET}   
  H1 Collab., I. Abt et al., DESY  93-103 (1993).    
\bibitem{ZEUSDET} 
  ZEUS Collab., M. Derrick et al., Phys. Lett. {\bf B293} (1992) 465.   
\bibitem{H1F295}  
  H1 Collab., contributed paper to 1995 EPS conference, Bruessel,       
  EPS-470 (1995). 
\bibitem{ZEUSF295}        
  ZEUS Collab., contributed paper to 1995 EPS conference, Bruessel,     
  EPS-392 (1995). 
\bibitem{H1RAD}   
  H1 Collab.,  contributed paper to 1995 EPS conference, Bruessel,      
  EPS-472 (1995). 
\bibitem{DJANGO}  
  G. A. Schuler and H. Spiesberger, Proceedings of the Workshop Physics at      
  HERA, vol. 3, eds. W. Buchm\"uller, G. Ingelman, DESY (1992) 1419.    
\bibitem{HERACLES}        
  A. Kwiatkowski, H. Spiesberger and H.-J. M\"ohring,   
  Computer Phys. Comm. {\bf 69} (1992) 155.  
\bibitem{LEPTO}   
  G. Ingelman, Proceedings of the Workshop Physics at HERA,     
  vol. 3, eds. W. Buchm\"uller, G. Ingelman, DESY (1992) 1366.  
\bibitem{MRSH}    
  A. D. Martin, W. J. Stirling and  R. G. Roberts, Proceedings of the   
  Workshop on Quantum Field Theory Theoretical Aspects of High Energy   
  Physics, eds. B. Geyer and E. M. Ilgenfritz (1993) 11.        
\bibitem{CDM}     
  L. L\"onnblad, Computer Phys. Comm. {\bf 71} (1992) 15.       
\bibitem{H1FLOW}  
  H1 Collab., I. Abt et al., Z. Phys. {\bf C63} (1994) 377.     
\bibitem{ZEUSFLOW}        
  ZEUS Collab., M. Derrick et al., Z. Phys. {\bf C59} (1993), 231.      
\bibitem{HERWIG}  
  G. Marchesini et al., Computer Phys. Comm. {\bf 67} (1992) 465.       
\bibitem{GEANT} R. Brun et al., GEANT3 User's Guide, 
  CERN--DD/EE 84--1, Geneva (1987).  
\bibitem{ALTMAR}  
  G. Altarelli and G. Martinelli, Phys. Lett. {\bf B76} (1978) 89.      
\bibitem{H1DIFF}  
  H1 Collab.,  contributed paper to 1995 EPS conference, Bruessel,      
        EPS-0491 (1995).  
\bibitem{ZEUSDIFF}        
  ZEUS Collab.,  contributed paper to 1995 EPS conference, Bruessel,    
  EPS-0393 (1995).        
\bibitem{H1QCD}   
  H1 Collab., S. Aid et al., Phys. Lett. {\bf B354} 494 (1995).         
\bibitem{NMC}     
  NMC Collab., P. Amaudruz et al., Phys. Lett. {\bf B259} (1992) 159.   
\bibitem{BCDMS}   
  BCDMS Collab., A. C. Benvenuti et al., Phys. Lett. {\bf B237} (1990)  
  592.    
\bibitem{BALL}    
  R.D. Ball, S. Forte, Phys. Lett. {\bf B335} (1994) 77.        
\bibitem{GRV}     
  M. Gluck, E. Reya and A. Vogt, Z. Phys. {\bf C67} (1995) 433.         
\bibitem{E665}    
  E665 Collab., M.R. Adams et al., Phys. Rev. Lett. {\bf 75} (1995) 1466.       
\bibitem{H1STOT}  
  H1 Collab., S. Aid et al., DESY 95-162 (1995).     
\bibitem{ZEUSSTOT}        
  ZEUS Collab., M. Derrick et al., Z. Phys. {\bf C63} (1994) 391.       
\bibitem{DOLA}    
  A. Donnachie and P.~V. Landshoff, Z. Phys. {\bf C61} (1994) 139.      
\bibitem{CTKM}    
  A. Capella et al., Phys. Lett. {\bf B337} (1994) 358.         
\bibitem{CTEQ}    
  CTEQ Collab., J. Botts et al., Phys. Lett. {\bf304B} (1993) 15; \\    
  CTEQ Collab., J. Botts et al.  ( to be published). 
\bibitem{MRSA}    
  A.D. Martin, W.J. Stirling and R.G. Roberts, RAL preprint RAL-95-021  
  (1995).         
\end{thebibliography}
\end{document}